\documentclass[12pt,english]{paper}
\usepackage[T1]{fontenc}
\usepackage[latin1]{inputenc}
\usepackage{graphicx}

\makeatletter

\makeatletter

\usepackage{lineno}

\makeatletter

\usepackage{geometry}

\makeatletter

\usepackage{epsfig}

\makeatother

\makeatother

\makeatother

\usepackage{babel}
\makeatother
\begin{document}

\title{Improving Application of Bayesian Neural Networks to Discriminate Neutrino Events from Backgrounds in
Reactor Neutrino Experiments}

\author{Ye Xu$^a$%
\thanks{Corresponding author, e-mail address: xuye76@nankai.edu.cn%
}, WeiWei Xu$^a$, YiXiong Meng$^a$, Bin Wu$^a$}

\maketitle
\begin{flushleft}
$^a$Department of Physics, Nankai University, Tianjin 300071, China
\par\end{flushleft}

\begin{abstract}
The application of Bayesian Neural Networks(BNN) to discriminate
neutrino events from backgrounds in reactor neutrino experiments has
been described in Ref.\cite{key-1}. In the paper, BNN are also used
to identify neutrino events in reactor neutrino experiments, but the
numbers of photoelectrons received by PMTs are used as inputs to BNN
in the paper, not the reconstructed energy and position of events.
The samples of neutrino events and three major backgrounds from the
Monte-Carlo simulation of a toy detector are generated in the signal
region. Compared to the BNN method in Ref.\cite{key-1}, more
$^{8}$He/$^{9}$Li background and uncorrelated background in the
signal region can be rejected by the BNN method in the paper, but
more fast neutron background events in the signal region are
unidentified using the BNN method in the paper. The uncorrelated
background to signal ratio and the $^{8}$He/$^{9}$Li background to
signal ratio are significantly improved using the BNN method in the
paper in comparison with the BNN method in Ref.\cite{key-1}. But the
fast neutron background to signal ratio in the signal region is a
bit larger than the one in Ref.\cite{key-1}.
\end{abstract}
\begin{flushleft}
\par\end{flushleft}

\section{Introduction}

The main goals of reactor neutrino experiments are to detect $\bar{\nu_{e}}\rightarrow\bar{\nu_{x}}$
oscillation and precisely measure the mixing angle of neutrino oscillation
$\theta_{13}$. The experiment is designed to detect reactor $\bar{\nu_{e}}$'s
via the inverse $\beta$-decay reaction

\begin{center}
$\bar{\nu_{e}}+p\rightarrow e^{+}+n$.
\par\end{center}

\begin{flushleft}
The signature is a delayed coincidence between $e^{+}$ and the
neutron captured signals. In the paper, only three important sources
of backgrounds are taken into account and they are the uncorrelated
background from natural radioactivity and the correlated backgrounds
from fast neutrons and $^{8}$He/$^{9}$Li. The backgrounds like the
neutrino events consist of two signals, a fast signal and a delay
signal. It is vital to separate neutrino events from backgrounds
accurately in the reactor neutrino experiments.  Bayesian neural
networks (BNN)\cite{key-2} are algorithms of the neural networks
trained by Bayesian statistics. They are not only non-linear
functions as neural networks, but also controls model complexity. So
their flexibility makes them possible to discover more general
relationships in data than traditional statistical methods and their
preferring simple models make them possible to solve the
over-fitting problem better than the general neural
networks\cite{key-3}.  BNN have been used to particle identification
and event reconstruction in the experiments of the high energy
physics, such as Ref.\cite{key-1,key-4,key-5,key-6}. The application
of BNN to discriminate neutrino events from backgrounds in reactor
neutrino experiments has been described in Ref.\cite{key-1}. In the
paper, BNN are also used to identify neutrino events in the signal
region\cite{key-1} in reactor neutrino experiments, but the numbers
of photoelectron received by PMTs are used as inputs to BNN, not the
reconstructed energy and position of events.
\par\end{flushleft}

\section{The Classification with BNN\cite{key-1,key-2,key-6}}

The idea of BNN is to regard the process of training a neural
network as a Bayesian inference. Bayes' theorem is used to assign a
posterior density to each point, $\bar{\theta}$, in the parameter
space of the neural networks. Each point $\bar{\theta}$ denotes a
neural network. In the methods of BNN, one performs a weighted
average over all points in the parameter space of the neural
network, that is, all neural networks. The methods are described in
detail in Ref.\cite{key-1,key-2,key-6}. The posterior density
assigned to the point $\bar{\theta}$, that is, to a neural network,
is given by Bayes' theorem

\begin{center}
\begin{equation}
p\left(\bar{\theta}\mid x,t\right)\propto\mathit{p\left(t\mid
x,\bar{\theta}\right)p\left(\bar{\theta}\right)}\end{equation}

\par\end{center}
Where $x$ is a set of input data which corresponds to a set of
target $t$. The likelihood $p\left(t\mid x,\bar{\theta}\right)$ can
be obtained by using a training sample. And a Gaussian prior is
specified for each weight using
the Bayesian neural networks package of Radford Neal%
\footnote{R. M. Neal, \emph{Software for Flexible Bayesian Modeling and Markov
Chain Sampling}, http://www.cs.utoronto.ca/\textasciitilde{}radford/fbm.software.html%
}. Given an event with data $x'$, an estimate of the probability
that it belongs to the signal is given by the weighted average

\begin{center}
\begin{equation}
\bar{y}\left(x'|x,t\right)=\int
y\left(x',\bar{\theta}\right)p\left(\bar{\theta}\mid
x,t\right)d\bar{\theta}\end{equation}

\par\end{center}
\begin{flushleft}
Currently, the only way to perform the high dimensional integral in
Eq. (2) is to sample the density $p\left(\bar{\theta}\mid
x,t\right)$ with Markov Chain Marlo Carlo (MCMC)
methods\cite{key-2,key-7,key-8,key-9}. In MCMC methods, one steps
through the $\bar{\theta}$ parameter space in such a way that points
are visited with a probability proportional to the posterior
density, $p\left(\bar{\theta}\mid x,t\right)$. Points where
$p\left(\bar{\theta}\mid x,t\right)$ is large will be visited more
often than points where $p\left(\bar{\theta}\mid x,t\right)$ is
small.
\par\end{flushleft}

\begin{flushleft}
Eq. (2) approximates the integral using the average
\par\end{flushleft}

\begin{center}
\begin{equation}
\bar{y}\left(x'\mid x,t\right)\approx\frac{1}{L}\sum_{i=1}^{L}y\left(x',\bar{\theta_{i}}\right)\end{equation}

\par\end{center}

\begin{flushleft}
where $L$ is the number of points $\bar{\theta}$ sampled from $p\left(\bar{\theta}\mid x,t\right)$.
Each point $\bar{\theta}$ corresponds to a different neural network
with the same structure. So the average is an average over neural
networks, and the probability of the data $x'$ belongs to the signal.
The average is closer to the real value of $\bar{y}\left(x'\mid x,t\right)$,
when $L$ is sufficiently large.
\par\end{flushleft}

\section{Toy Detector and Monte-Carlo Simulation\cite{key-5}}

In the paper, a toy detector is used to simulate central detectors
in the reactor neutrino experiments, such as Daya Bay
experiment\cite{key-10} and Double Chooz experiment\cite{key-11},
with CERN GEANT4 package\cite{key-12}. The toy detector is the same
as Ref.\cite{key-5}. A total of 366 PMTs are arranged in 8 rings of
30 PMTs on the lateral surface of the oil region, and in 5 rings of
24, 18, 12, 6, 3 PMTs on the top and bottom caps.
\par
The responses of neutrino events and backgrounds deposited in the
toy detector are simulated with GEANT4. Although the physical
properties of the scintillator and the oil (their optical
attenuation length, refractive index and so on) are wave-length
dependent, only averages\cite{key-13} (such as the optical
attenuation length of Gd-LS with a uniform value is 8 meter and the
one of LS is 20 meter) are used in the detector simulation.

\begin{flushleft}
According to the anti-neutrino interaction in detectors of the
reactor neutrino experiments\cite{key-14}, a neutrino event is
uniformly generated throughout Gd-LS region (see Fig. 1). A
uncorrelated background event is generated in such a way that a
$\gamma$ event generated on the base of the energy distribute of the
natural radioactivity in the proposal of the Day Bay
experiment\cite{key-10} is regarded as the fast signal, a neutron
event of the single signal is regarded as the delay signal, its
delay time is uniformly generated from 2 $\mu$s to 100 $\mu$s and
the positions of the fast signal and the delay signal are uniformly
generated throughout Gd-LS region. A fast neutron event is uniformly
generated throughout Gd-LS region and its energy are uniformly
generated from 0 MeV to 50 MeV, therein an event of two signals are
regarded as a fast neutron background event. Since the behaviors of
$^{8}$He/$^{9}$Li decay events in detectors couldn't be simulated by
the Geant4 package, a $^{8}$He/$^{9}$Li event is generated in such a
way that the neutron signal from a fast neutron event is regarded as
its delay signal, an electron event generated at the same position
as the fast neutron event on the base of the energy distribute of
$^{8}$He/$^{9}$Li events in the proposal of the Day Bay
experiment\cite{key-10} is regarded as its fast signal in the paper.
\par\end{flushleft}

Energies and positions of neutrino events and backgrounds are
reconstructed by the method in Ref.\cite{key-5}. The signal region
is determined by using the reconstructed energies and positions, as
well as the neutron delay time(described in Ref.\cite{key-1}).

\section{Neutrino Discrimination with BNN}
Choosing inputs to BNN is vital to identify neutrino events . The
reconstructed energies, the distance between reconstructed the
positions of neutron and positron and the neutron delay time were
used as inputs to the BNN method in Ref.\cite{key-1}, but the
energies and the distance are both the reconstructed physics
variables, and they make BNN discriminations worse because of their
reconstruction uncertainties. So we try to use raw data as inputs to
BNN. Obviously, the numbers of photoelectrons received by 366 PMTs
are rawer than the reconstructed variables. An event consists of two
signals (a fast signal and a delay signal), so if the numbers of
photoelectron received by PMTs will be directly used as inputs to
BNN, BNN will have 732 inputs at least. It will take too much time
to run a BNN program in a general computer because of such many
inputs. The method of reducing inputs to BNN in the paper is that
the photoelectrons received by several neighboring PMTs are added
up. That is several neighboring PMTs incorporate a PMT patch. In the
paper, a PMT patch is a 3(azimuth direction)$\times$4(z direction)
PMTs array on the detector lateral surface or a 120$^\circ$
sector(including 21 PMTs) on the detector top and bottom caps. The
delay time between two signals is very important to discriminate
neutrino events from the uncorrelated background, so the number of
photoelectrons received by a patch is multiply by the delay time,
and the result is used as the inputs to all neural networks, which
have the same structure. Then all the networks have a input layer of
52 inputs, the single hidden layer of fifteen nodes and a output
layer of a single output which is just the probability that an event
belongs to the neutrino event. Discriminating neutrino events from
backgrounds is actually a binary response problem, that is the
target is '1' or '0'. Neutrino events are labeled by t=1, and
background events are labeled by t=0. So the output of BNN has to be
a number between 0 and 1. If the output is less than 0.5, the event
is regarded as a background event, and If the output is larger than
0.5, the event is regarded as a neutrino event.
\par
 A Markov chain of neural networks is generated using
the Bayesian neural networks package of Radford Neal, with a
training sample consisting of neutrino events and background events.
One thousand iterations, of twenty MCMC steps each, are used. The
neural network parameters are stored after each iteration, since the
correlation between adjacent steps is very high. That is, the points
in neural network parameter space are saved to lessen the
correlation after twenty steps here. It is also necessary to discard
the initial part of the Markov chain because the correlation between
the initial point of the chain and the point of the part is very
high. The initial three hundred iterations are discarded here. It
takes about 120 hours to run 1000 iterations on a computer with two
3.4GHz Intel Pentium D processors (only one of which are used).
\par
Neutrino identification efficiencies are defined by the ratios
between the number of the events in neutrino test sample regarded as
neutrinos and the number of neutrino test sample. Background
identification efficiencies are defined by the ratios between the
numbers of the events in background test samples regarded as
background events and the numbers of background test samples. The
identification efficiencies are measured with the test sample which
is different from the training sample. Other 3000 events each of the
neutrino and the three backgrounds are used to test the
identification capability of the trained BNN. In the paper, BNN are
trained by the different training samples, which consist of
neutrinos and three backgrounds at different rates, since the
different identification efficiencies are obtained using those BNN.

\section{Results and Discussion}
 As Tab. 1 shows, most neutrino events,
uncorrelated background events and $^{8}$He/$^{9}$Li background
events in the signal region can be identified using the BNN method
in the paper, but only a small part of fast neutron background
events can be identified using the BNN method in the paper. Since
most fast neutron events can't be discriminate from neutrino events
using the BNN method in the paper, neutrino discriminations are
concerned with neutrino rates in training samples, as well as ratios
of neutrino events and fast neutron events in training samples. The
neutrino discrimination in the signal region increases from 90.5\%
to 93.7\% with the increase of the neutrino rate from 50.0\% to
57.1\% in the training sample using the BNN method in the paper. And
the neutrino discrimination also increases from 90.5\% to 94.1\%
with the increase of the ratio of neutrino events and fast neutron
events from 2:1 to 3:1 in the training sample. The different
background to signal ratios in the signal region are obtained using
the BNN trained by the training samples consisting of neutrino
events and background events at different rates in the reactor
neutrino experiments.
\par
Neutrino events are discriminated from fast neutrons and
$^{8}$He/$^{9}$Li events via their fast signals identification, that
is positron signals from neutrino events are separated from recoil
proton signals from fast neutrons and electron signals from
$^{8}$He/$^{9}$Li events.  $\gamma$ signals induced by positrons and
recoil protons are closer to point sources, but $\gamma$ signals
induced by electrons are closer to line sources. There is an effect
on the distribution of photoelectrons over all the PMTs in the
detector due to the difference between a point source and a line
source. The effect can be extracted from the inputs by the BNN
method in the paper. So neutrino events can be better discriminated
from $^{8}$He/$^{9}$Li events using the BNN method in the paper, but
distinguishing between neutrinos events and fast neutrons becomes
worse using the BNN method in the paper.
\begin{flushleft}
The events in the signal region can be identified using BNN one by
one, once those BNN are trained by training samples. If the BNN
method in the paper is used to the reactor neutrino experiments, the
background to signal ratios will be changed. We only roughly
estimate the changes here. We assume that the uncorrelated
background fraction in the signal region is $A/N$, the fast neutrons
background fraction in the signal region is $F/N$, and the
$^{8}$He/$^{9}$Li background fraction in the signal region is $L/N$
in the reactor neutrino experiments. Those background fractions are
very low (for example, they are <0.2\%, 0.1\%, 0.3\% in one of the
near detector claimed by the proposal of the Daya Bay
experiment\cite{key-10}, respectively). If neutrino events are
discriminated from background events using the BNN method in
Ref.\cite{key-1}, the background to signal ratios can reach
0.2*(A/N), 0.68*(F/N) and 0.66*(L/N), respectively. If the
efficiencies of the first column in Tab. 1 are use to the
estimation, we get the result of the identification using the BNN
method in the paper:
\begin{center}
Uncorrelated background/Signal=(A/N)*(1-0.983)/0.941=0.018*(A/N)
\par
Fast neutrons Background/Signal=(F/N)*(1-0.293)/0.941=0.751*(F/N)
\par
$^{8}$He/$^{9}$Li
Background/Signal=(L/N)*(1-0.913)/0.941=0.092*(L/N)
\par\end{center}

As the above equations show, the uncorrelated background to signal
ratio and $^{8}$He/$^{9}$Li background to signal ratio in the signal
region are significantly improved using the BNN method in the paper
in comparison with the BNN method in Ref.\cite{key-1}. And the fast
neutron background to signal ratio is a bit larger than the one in
Ref.\cite{key-1}. But the fast neutron fraction in the signal region
is lower than the ones of the uncorrelated background and
$^{8}$He/$^{9}$Li background, so the total background to signal
ratio using the BNN method in the paper is much lower than the one
in Ref.\cite{key-1}. In a word, the BNN method in the paper can be
applied to discriminate neutrino events from background events
better than the BNN method in Ref.\cite{key-1} and the method based
on the cuts in reactor neutrino experiments.
\par\end{flushleft}

\section{Acknowledgements }

This work is supported in part by the National Natural Science
Foundation of China (NSFC) under the contract No. 10605014, the
national undergraduate innovative plan of China under the contract
No.081005517 and the physical base of Nankai University under the
contract No. J0730315.

\newpage{}

\begin{table}

\caption{The different identification efficiencies are obtained with
the BNNs trained by the different training samples, which consist of
the neutrino and three backgrounds at different rates. The term
after $\pm$ is the statistical error of the identification
efficiencies. The numbers of the train samples are 24000,
respectively. The 3000 events each of the uncorrelated background,
fast neutron and $^{8}$He/$^{9}$Li are regarded as the test sample.}
\begin{tabular}{|c|c|c|c|c|}\hline
 neutrino rate (\%)& 50.0& 50.0&
54.5& 57.1\\ uncorrelated background rate (\%)& 16.7&
12.5& 9.1& 9.5 \\ fast neutron rate (\%)& 16.7& 25.0& 27.3& 23.8\\
$^{8}$He/$^{9}$Li rate (\%) & 16.7& 12.5& 9.1& 9.5\\\hline
neutrino eff.(\%)& 94.1$\pm$0.43& 90.5$\pm$0.54& 92.6$\pm$0.48& 93.7$\pm$0.44\\
uncorrelated background eff.(\%)& 98.3$\pm$0.24& 98.1$\pm$0.25&
96.4$\pm$0.34& 96.7$\pm$0.33\\ fast neutrons eff.(\%)&
29.3$\pm$0.83& 35.8$\pm$0.88& 34.6$\pm$0.87&
32.8$\pm$0.86\\
 $^{8}$He/$^{9}$Li eff.(\%) &
91.3$\pm$0.51& 90.6$\pm$0.53& 87.7$\pm$0.60& 87.5$\pm$0.60\\\hline
\end{tabular}
\end{table}

\begin{figure}
\includegraphics[width=16cm,height=16cm]{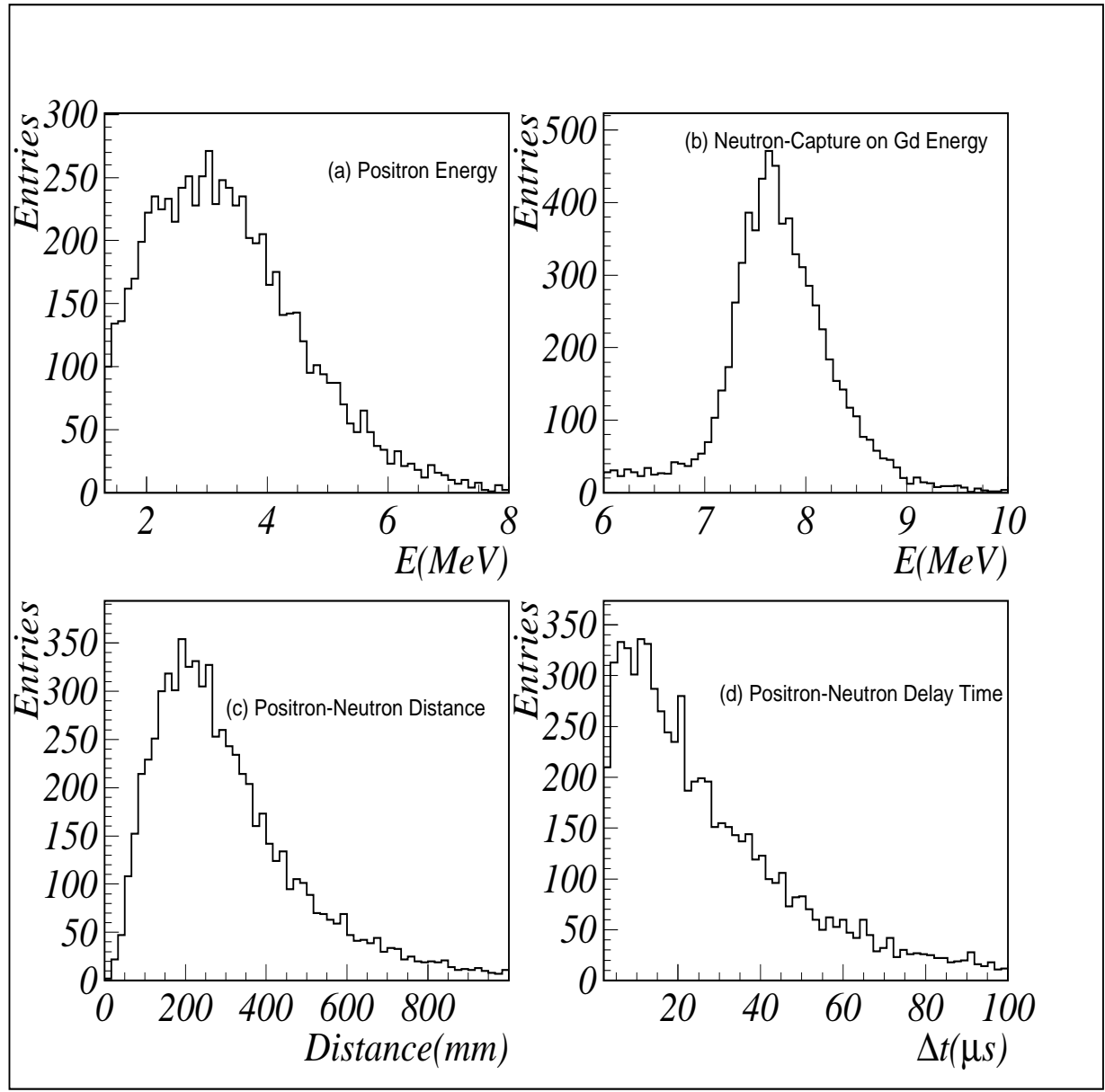}

\caption{The neutrino events for the Monte-Carlo simulation of the toy detector
are uniformly generated throughout Gd-LS region. (a) is the distribution
of the positron energy; (b) is the distribution of the energy of the
neutron captured by Gd; (c) is the distribution of the distance between
the positron and neutron positions; (d) is the distribution of the
delay time of the neutron signal.}
\end{figure}

\end{document}